\documentstyle[12pt,epsfig]{article}
\input epsf

\begin{document}

\thispagestyle{empty}
\renewcommand{\thefootnote}{\fnsymbol{footnote}}

\begin{flushright}
{\small
SLAC--PUB--7395\\
January 1997\\}
\end{flushright}

\vspace{.8cm}

\begin{center}
{\large \bf
Measurement of Leading Particle Effects in Decays of Z$^{0}$ Bosons into Light
Flavors\footnote{Work supported by Department of Energy contracts:
  DE-FG02- 91ER40676 (BU),
  DE-FG03- 91ER40618 (UCSB),
  DE-FG03- 92ER40689 (UCSC),
  DE-FG03- 93ER40788 (CSU),
  DE-FG02- 91ER40672 (Colorado),
  DE-FG02- 91ER40677 (Illinois),
  DE-AC03- 76SF00098 (LBL),
  DE-FG02- 92ER40715 (Massachusetts),
  DE-FC02- 94ER40818 (MIT),
  DE-FG03- 96ER40969 (Oregon),
  DE-AC03- 76SF00515 (SLAC),
  DE-FG05- 91ER40627 (Tennessee),
  DE-FG02- 95ER40896 (Wisconsin),
  DE-FG02- 92ER40704 (Yale);
  National Science Foundation grants:
  PHY-91- 13428 (UCSC),
  PHY-89- 21320 (Columbia),
  PHY-92- 04239 (Cincinnati),
  PHY-95- 10439 (Rutgers),
  PHY-88- 19316 (Vanderbilt),
  PHY-92- 03212 (Washington);
  The UK Particle Physics and Astronomy Research Council
  (Brunel and RAL);
  The Istituto Nazionale di Fisica Nucleare of Italy
  (Bologna, Ferrara, Frascati, Pisa, Padova, Perugia);
  The Japan-US Cooperative Research Project on High Energy Physics
  (Nagoya, Tohoku);
  The Korea Science and Engineering Foundation (Soongsil).}}

\vspace{1cm}

The SLD Collaboration$^{**}$\\
Stanford Linear Accelerator Center, Stanford University,
Stanford, CA  94309\\

\end{center}

\vfill

\begin{center}
{\large \bf
Abstract }
\end{center}

{\bf 
We present evidence for leading particle production in hadronic decays of the
$Z^{0}$ boson to light-flavor jets.  A polarized electron beam was used to tag
quark and antiquark jets, and a vertex detector was employed to reject
heavy-flavor events.  Charged hadrons were identified with a Cherenkov ring
imaging detector.  In the quark jets, more high-momentum p, $\Lambda$, $K^{-}$,
and $\overline{K}^{*0}$ were observed than their antiparticles, and vice versa
for antiquark jets, providing direct evidence that the higher-momentum
particles in jets are more likely to carry the primary quark or antiquark from
the $Z^{0}$ decay, and that $s\bar{s}$ production is suppressed in
fragmentation.
}
\vfill

\begin{center} 

{\it Submitted to Physical Review Letters}
\end{center}

\newpage



%
\pagestyle{plain}


%

A fundamental issue in strong-interaction jet fragmentation is that of the
transport of quantum numbers of primary interacting partons into the observed
final-state particles.  In non-diffractive hadron-hadron collisions,
final-state particles with large values of the longitudinal beam momentum
fraction $x_{F}$ have been observed that contain one or more valence quarks of
the same type as those in one or both of the initial-state particles.  This has
been interpreted in terms of an initial-state quark participating in the
collision and being carried in a particular ``leading'' final-state particle
that tends to have a large fraction of the energy of the resulting
jet~\cite{basile81}.  In $e^{+}e^{-}\rightarrow c\bar{c}$ ($b\bar{b}$) events,
$D\ (B)$ hadrons have been found~\cite{longlist} to carry a large fraction of
the beam energy and to be produced at a rate of approximately two per
$c\bar{c}$ or $b\bar{b}$ event, indicating that these hadrons are produced
predominantly as leading particles.

Such leading particle production in jet fragmentation is predicted by several
iterative models of the hadronization process~\cite{models}.  However, the
extent to which this effect is present in light-flavor jets ($u$, $\bar{u}$,
$d$, $\bar{d}$, $s$, or $\bar{s}$) in $e^{+}e^{-}$ interactions has not been
studied experimentally because of difficulties involved with tagging jets
initiated by a specific light flavor and with separating quark jets from
antiquark jets.  If such a separation were achieved, a signature for the
leading particle effect in a sample of quark jets would be an excess of a
hadron species containing the isolated valence quark type over its
antiparticle, and vice versa for antiquark jets.  One could then study the
momentum distributions, flavors and spin states of leading hadrons and
antihadrons in each such flavor sample.

If one could separate samples of light quark ($u$, $d$, $s$) and antiquark
($\bar{u}$, $\bar{d}$, $\bar{s}$) jets, then a leading particle effect might
appear as an excess in the quark sample of baryons over antibaryons, since the
valence constituents of baryons are quarks rather than antiquarks.  Also, the
cross sections for $e^{+}e^{-}\rightarrow u\bar{u}$ and $e^{+}e^{-}\rightarrow
d\bar{d}$ or $s\bar{s}$ are not in general equal, so a signal in quark jets for
leading production of charged mesons, such as $\pi^{-}$ and $K^{-}$, might be
visible.  Furthermore, one might observe an excess of a meson over its
antimeson if it is produced more often in jets initiated by one valence flavor
rather than the other.  For example, a suppression of $s\bar{s}$ relative to
$u\bar{u}$ and $d\bar{d}$ production in the fragmentation process might cause
more leading $K^{-}$ ($\overline{K}^{*0}$) to be produced in $s$ jets than in
$\bar{u}$ ($\bar{d}$) jets.

In this letter we present the first study of leading particle production in
light flavor jets in $e^{+}e^{-}$ annihilation, using 150,000 hadronic $Z^{0}$
decay events produced by the SLAC Linear Collider (SLC) and recorded in the SLC
Large Detector (SLD) from 1993 to 1995.  We define a particle to be leading if
it carries a primary quark or antiquark, namely the $q$ or $\bar{q}$ in
$e^{+}e^{-}\rightarrow Z^{0}\rightarrow q\bar{q}$, where $q = u, d,$ or $s$. 
We separated jets initiated by primary quarks from those initiated by primary
antiquarks by utilizing the electroweak forward-backward production asymmetry
in the polar angle, enhanced by the high SLC electron beam polarization.  We
suppressed the large background from heavy-flavor ($Z^{0}\rightarrow c\bar{c}$
or $b\bar{b}$) events, in which the decay products of a heavy hadron can
exhibit a leading particle effect, by using information from the Vertex
Detector (VXD)~\cite{vxd}.  The Cherenkov Ring Imaging Detector
(CRID)~\cite{crid} was used to identify charged hadrons.  We measured the
production rates of $\pi^{-}$, $K^{-}$, $\overline{K}^{*0}$, p, and $\Lambda$
as a function of momentum in light-quark jets and compared them with the rates
of their respective antiparticles.  We interpret the observed differences in
terms of leading particles.


A description of the detector, trigger, track and hadronic event selection, and
Monte Carlo simulation is given in Ref.~\cite{impact}.  Cuts were applied in
order to select events well-contained within the detector acceptance, resulting
in a sample of approximately 90,000 events.  Heavy-flavor events often include
tracks associated with separated decay points of short-lived heavy hadrons, and
were suppressed by requiring all tracks passing a set of quality cuts to
extrapolate to within three standard deviations from the interaction point in
the plane transverse to the beam.  The selected sample was estimated from our
Monte Carlo simulation to consist of 85\% light-flavor events, with residual
backgrounds of 12\% $c\bar{c}$ and 3\% $b\bar{b}$ events.

$Z^{0}$ bosons decay predominantly into a left-handed quark and a right-handed
antiquark.  In $e^{+}e^{-}\rightarrow Z^{0}\rightarrow q\bar{q}$ events, when
the electron beam has longitudinal polarization $P_{e}$, the quark prefers to
follow the electron (positron) beam direction for left-(right-) handed $e^{-}$
beam, and its polar angle $\theta$ with respect to the electron beam is
distributed as $(1 + \cos^{2} \theta + 2A_{q}A_{Z}\cos\theta )$, where $A_{Z} =
(A_{e}-P_{e})/(1-A_{e}P_{e})$, and $A_{e}$ and $A_{q}$ are the asymmetry
parameters for electrons and quarks respectively.  In the Standard Model $A_{e}
= 0.16$, $A_{u} = A_{c} = 0.67$ and $A_{d} = A_{s} = A_{b} = 0.94$.  For this
analysis we considered all events to consist of one jet in each of the two
hemispheres separated by the plane perpendicular to the thrust
axis~\cite{thrust}, and required the thrust axis polar angle $\theta_t$ to
satisfy $|\cos\theta_t | > 0.2$.  Defining the forward direction to be along
the electron beam, the quark jet was defined to comprise the set of tracks in
the forward (backward) hemisphere for events recorded with left-(right-) handed
electron beam.  The opposite jet in each event was defined to be the antiquark
jet.  For roughly two-thirds of the sample, $|P_{e}| = 0.77$~\cite{polnew}, for
the remainder $|P_{e}| = 0.63$~\cite{polold}, and there were equal numbers of
left- and right-handed beam pulses.  For these conditions, the Standard Model
at tree level predicts the purities of the quark- and antiquark-tagged samples
to be about 73\%.


We then measured the production rates per light quark jet
\begin{eqnarray}
R^{q}_{h} &=& {1\over{2N_{evts}}}{d\over{dx_{p}}}\left[ N(q\rightarrow
h)+N(\bar{q}\rightarrow\bar{h})\right],\\
R^{q}_{\bar{h}} &=& {1\over{2N_{evts}}}{d\over{dx_{p}}}\left[
N(q\rightarrow\bar{h})+N(\bar{q}\rightarrow h)\right],
\end{eqnarray}
where: $q$ and $\bar{q}$ represent light-flavor quark and antiquark jets
respectively; $N_{evts}$ is the total number of events in the sample; $h$
represents any of the identified hadrons $\pi^{-}$, $K^{-}$,
$\overline{K}^{*0}$, p, and $\Lambda$, and $\bar{h}$ indicates the
corresponding antiparticle; $x_{p}$ is the scaled momentum $2p/\sqrt{s}$ of the
hadron, where $p$ is its magnitude of momentum and $\sqrt{s}$ is the
$e^{+}e^{-}$ center-of-mass energy.  Then, for example, $N(q\rightarrow h)$ is
the number of hadrons of type $h$ in light quark jets.

The identification of $\pi^{\pm}$, $K^{\pm}$, p, and $\bar{\rm p}$ was achieved
by reconstructing emission angles of individual Cherenkov photons radiated by
charged particles passing through liquid and gas radiator systems of the SLD
CRID.  For each track, a likelihood was constructed for each of these particle
hypotheses, based upon the number of detected photons and their measured
angles, and the expected number of photons, Cherenkov angle, and background. 
Particle separation was based on the differences among the likelihoods. 
Identification was achieved~\cite{partid,pavel} over the momentum range
$0.5<p<35\ GeV/c$.

Positively charged tracks in the quark-tagged sample and negatively charged
tracks in the antiquark-tagged sample gave consistent results and were combined
into one sample.  In each $x_{p}$ bin, identified $\pi$, $K$, and p were
counted, and these counts were unfolded using the inverse of the identification
efficiency matrix ${\bf E}$~\cite{partid,pavel}, and corrected for track
reconstruction efficiency to yield values of $R^{q}_{\pi^{+}}$,
$R^{q}_{K^{+}}$, and $R^{q}_{\rm p}$ in the tagged samples.  The same procedure
applied to the remaining tracks yielded $R^{q}_{\pi^{-}}$, $R^{q}_{K^{-}}$, and
$R^{q}_{\bar{\rm p}}$.  The elements $E_{ij}$, denoting the momentum-dependent
probability to identify a true $i$-type particle as a $j$-type particle, were
measured from the data for $i=\pi ,{\rm p}$ and $j=\pi ,K,{\rm p}$ using tracks
from selected $K^{0}_{S}$, $\tau$ and $\Lambda$ decays.  A detailed Monte Carlo
simulation was used to derive the remaining elements in terms of these measured
ones.

Candidate $\Lambda\rightarrow{\rm p}\pi^{-}$ and
$\bar{\Lambda}\rightarrow\bar{\rm p}\pi^{+}$ decays were selected by
considering all pairs of oppositely charged tracks that were inconsistent with
originating at the interaction point and passed a set of cuts~\cite{baird} on
vertex quality and flight distance.  Backgrounds from $K^{0}_{S}$ decays and
photon conversions were suppressed by using kinematic cuts.  Candidate
$K^{*0}\rightarrow K^+\pi^-$ and $\overline{K}^{*0}\rightarrow K^-\pi^+$ decays
were selected by considering all pairs of oppositely-charged tracks if one
track was identified in the CRID as a charged kaon, the other was not so
identified, and the tracks were consistent with intersecting at the interaction
point~\cite{dima}.

The $\Lambda$ candidates in quark-tagged jets and the $\bar{\Lambda}$
candidates in antiquark-tagged jets were assigned to one sample, and the
remaining $\Lambda /\bar{\Lambda}$ candidates to a second sample.  In each
$x_{p}$ bin, the number of observed $\Lambda /\bar{\Lambda}$ in each sample
was determined from a fit to the $p\pi$ invariant mass distribution.  These
signals were corrected for reconstruction efficiency to yield $R^{q}_{\Lambda}$
and $R^{q}_{\bar{\Lambda}}$ in the tagged samples.  The $\overline{K}^{*0}$ and
$K^{*0}$ candidates were similarly divided into two samples, and the $K\pi$
invariant mass distributions were fitted to obtain $R^{q}_{\overline{K}^{*0}}$
and $R^{q}_{K^{*0}}$.


In every $x_{p}$ bin, each measured $R^{q}_{h}$ and $R^{q}_{\bar{h}}$ was
further corrected for the contribution from residual heavy-flavor events,
estimated from our Monte Carlo simulation.  Finally, the corrected $R^{q}_{h}$
and $R^{q}_{\bar{h}}$ were unfolded for the purity of the quark jet tag.

\begin{figure}[p]
  \vspace*{0.5cm}
  \begin{center}
      \epsfxsize=0.50\textwidth
      \epsfig{figure=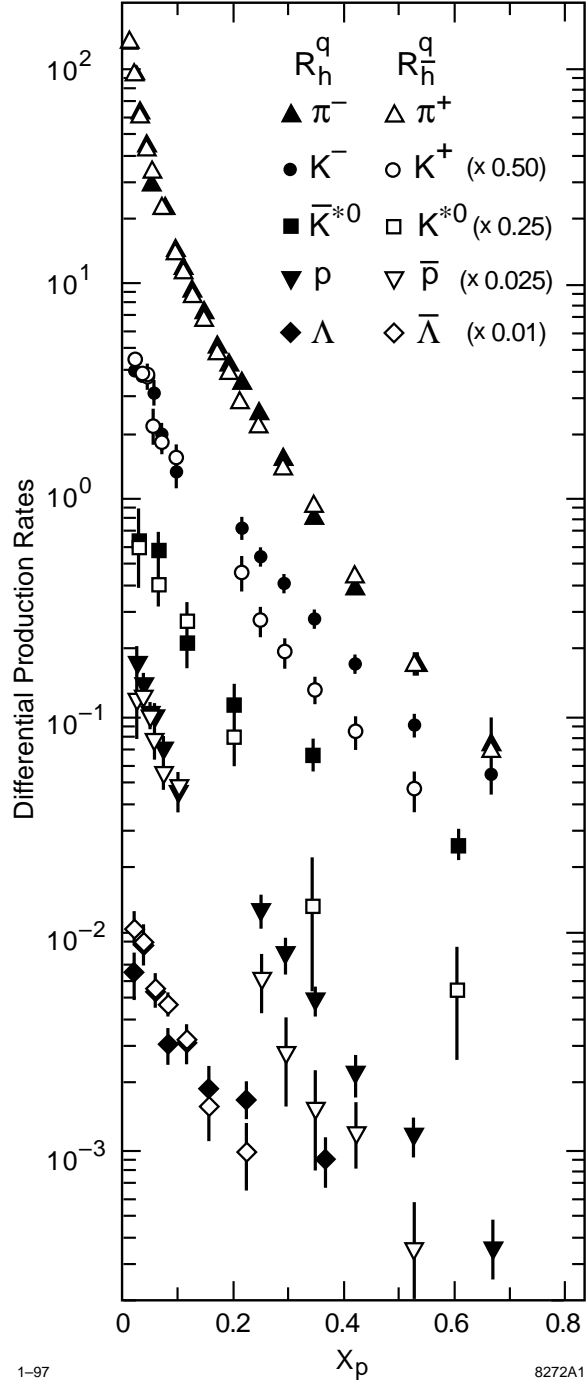}
     \caption{\label{rates}
      Differential production rates as a function of scaled momentum. 
      The ordinates represent average
      multiplicities per light quark jet per unit interval in scaled momentum.}
  \end{center}
\end{figure}

These production rates are shown in Figure~\ref{rates}.  There are no $K^{\pm}$
or p/$\bar{\rm p}$ points in the range $0.12 < x_p < 0.20$ due to the lack of
CRID particle separation in this region.  Systematic errors arising from the
uncertainties in the backgrounds in the identified particle samples, in the
measured electron beam polarization, and in the backgrounds from heavy-flavor
events were included and were found to be much smaller than the statistical
errors.  Not shown in the figure are uncertainties common to particles and
their respective antiparticles, including those arising from track
reconstruction and particle-identification efficiency.  These are typically
2-5\%.

\begin{figure}[p]
  \vspace*{0.5cm}
  \begin{center}
      \epsfxsize=0.50\textwidth
      \epsfig{figure=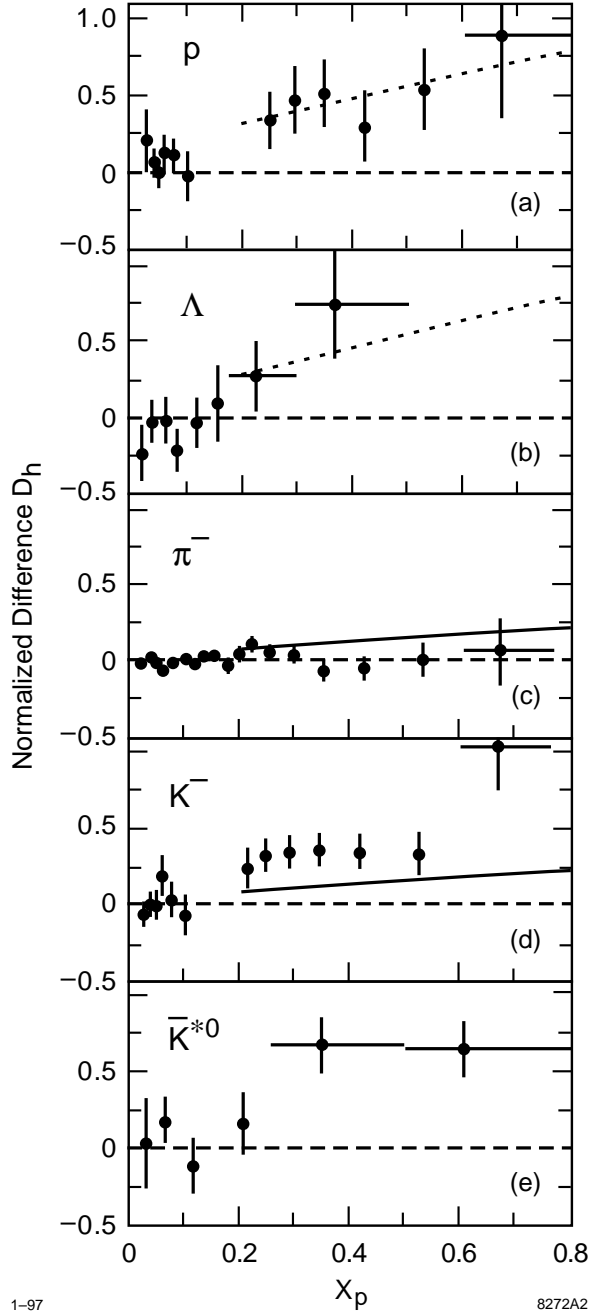}
    \caption{\label{diffs}
    Normalized production differences (dots) as a function of scaled momentum.
The horizontal error bars on selected points
indicate their bin widths.  The dotted lines represent a linear fit to the
$D_{p}$ and $D_{\Lambda}$ points for $x_p > 0.2$, and the solid lines are this
fit
scaled by the factor 0.27 discussed in the text.}
  \end{center}
\end{figure}

We define the difference between each particle and antiparticle production
rate, normalized by the sum: $$D_{h} =  {R^{q}_{h} - R^{q}_{\overline{h}}\over
R^{q}_{h} + R^{q}_{\overline{h}}},$$ for which the common systematic
uncertainties cancel.  As shown in Figure~\ref{diffs}, for each hadron $h$,
$D_{h}$ is consistent with zero for $x_p < 0.1$.  $D_{\pi^{-}}$ is also
consistent with zero for $x_p > 0.1$, but for the other hadrons $D_{h} > 0$ for
$x_p$ \raisebox{-.7ex}{$\stackrel{\textstyle >}{\sim}$} 0.2.  The JETSET
7.4~\cite{jetset} and HERWIG 5.8~\cite{herwig} fragmentation models were found
to reproduce these features qualitatively.

Since baryons contain no constituent antiquarks, we interpret the positive
$D_{p}$ and $D_{\Lambda}$ as evidence for leading baryon production in
light-flavor jets.  If pions and kaons exhibited similar leading effects, then
one would expect $D_{\pi^-} \approx D_{K^-}  \approx  0.27 D_{baryon}$, and
$D_{\overline{K}^{*0}} = 0$, assuming Standard Model quark couplings to the
$Z^0$.  For purposes of illustration, the result of a linear fit to the $D_{p}$
and $D_{\Lambda}$ points above $x_p=0.2$ was scaled by 0.27 and is shown in
Figs. 2(c) and 2(d).  The observed $D_{\pi^-}$ are below this line, and are
consistent with zero at all $x_p$, suggesting that either there is little
production of leading pions, or there is substantial background from
non-leading pions or pions from decays of resonances such as the $\rho$ and
$K^*$. 
For $x_p > 0.2$, we observe $D_{K^-} > 0.27 D_{baryon}$ and
$D_{\overline{K}^{*0}} > 0$.  This indicates both substantial production of
leading $K$ and $K^*$ mesons at high momentum, and a depletion of leading kaon
production in $u\bar{u}$ and $d\bar{d}$ events relative to $s\bar{s}$ events. 

Assuming these high-momentum kaons to be directly produced in the fragmentation
process, this amounts to a direct observation of a suppression of $s\bar{s}$
production from the vacuum with respect to $u\bar{u}$ or $d\bar{d}$ production. 
In the case of $K^{*0}$ mesons it has been suggested~\cite{lafferty} that this
effect can be used to measure the ``strangeness suppression parameter''
$\gamma_s$, that is an important component of models of hadronization, see e.g.
Ref.~\cite{jetset}.  Assuming {\it all} $K^{*0}$ and $\overline{K}^{*0}$ in the
range $x_{p}>0.5$ to be leading, we calculate $\gamma_s = 0.26\pm 0.12$,
consistent with values~\cite{saxon} derived from inclusive measurements of the
relative production rates of strange and non-strange, pseudoscalar and vector
mesons. 


In summary, we have studied leading particle effects in hadronic $Z^{0}$
decays.  In the light quark jets, we observed an excess of $\Lambda$ over
$\bar{\Lambda}$, and an excess of p over $\bar{\rm p}$. These differences
increase with momentum, and provide direct evidence for the ``leading
particle'' hypothesis that faster baryons are more likely to contain the
primary quark. No such difference was observed between $\pi^-$ and $\pi^+$
production. For kaons, we observed a significant excess of high momentum $K^-$
over $K^+$, and $\overline{K}^{*0}$ over $K^{*0}$, indicating that a fast kaon
is likely to contain a primary quark or antiquark from the $Z^0$ decay, and
that leading kaons are produced predominantly in $s\bar{s}$ events rather than
$d\bar{d}$ or $u\bar{u}$ events.


We thank the personnel of the SLAC accelerator department and the
technical
staffs of our collaborating institutions for their outstanding efforts
on our behalf. 


 

\newpage
\section*{$^{**}$List of Authors} 


\begin{center}
%
%
%
  \def\iADEL{$^{(1)}$}
  \def\iBOL{$^{(2)}$}
  \def\iBU{$^{(3)}$}
  \def\iBRUN{$^{(4)}$}
  \def\iUCSB{$^{(5)}$}
  \def\iUCSC{$^{(6)}$}
  \def\iCIN{$^{(7)}$}
  \def\iCSU{$^{(8)}$}
  \def\iCOLO{$^{(9)}$}
  \def\iCOL{$^{(10)}$}
  \def\iFER{$^{(11)}$}
  \def\iFRA{$^{(12)}$}
  \def\iILL{$^{(13)}$}
  \def\iLBL{$^{(14)}$}
  \def\iMIT{$^{(15)}$}
  \def\iMASS{$^{(16)}$}
  \def\iMISS{$^{(17)}$}
  \def\iMOSC{$^{(18)}$}
  \def\iNAG{$^{(19)}$}
  \def\iOREG{$^{(20)}$}
  \def\iPAD{$^{(21)}$}
  \def\iPERU{$^{(22)}$}
  \def\iPISA{$^{(23)}$}
  \def\iRUT{$^{(24)}$}
  \def\iRAL{$^{(25)}$}
  \def\iSOGANG{$^{(26)}$}
  \def\iSOONG{$^{(27)}$}
  \def\iSLAC{$^{(28)}$}
  \def\iTENN{$^{(29)}$}
  \def\iTOH{$^{(30)}$}
  \def\iVAND{$^{(31)}$}
  \def\iWASH{$^{(32)}$}
  \def\iWISC{$^{(33)}$}
  \def\iYALE{$^{(34)}$}
  \def\dead{$^{\dag}$}
  \def\andgen{$^{(a)}$}
  \def\andper{$^{(b)}$}
%
%
\mbox{K. Abe                 \unskip,\iNAG}
\mbox{K. Abe                 \unskip,\iTOH}
\mbox{T. Akagi               \unskip,\iSLAC}
\mbox{N.J. Allen             \unskip,\iBRUN}
\mbox{W.W. Ash               \unskip,\iSLAC$^\dagger$}
\mbox{D. Aston               \unskip,\iSLAC}
\mbox{K.G. Baird             \unskip,\iRUT}
\mbox{C. Baltay              \unskip,\iYALE}
\mbox{H.R. Band              \unskip,\iWISC}
\mbox{M.B. Barakat           \unskip,\iYALE
% ...
}
\it
  \vskip \baselineskip                   
%
%
%
  \iADEL
     Adelphi University,
     Garden City, New York 11530 \break
  \iBOL
     INFN Sezione di Bologna,
     I-40126 Bologna, Italy \break
  \iBU
     Boston University,
     Boston, Massachusetts 02215 \break
  \iBRUN
     Brunel University,
     Uxbridge, Middlesex UB8 3PH, United Kingdom \break
  \iUCSB
     University of California at Santa Barbara,
     Santa Barbara, California 93106 \break
  \iUCSC
     University of California at Santa Cruz,
     Santa Cruz, California 95064 \break
  \iCIN
     University of Cincinnati,
     Cincinnati, Ohio 45221 \break
  \iCSU
     Colorado State University,
     Fort Collins, Colorado 80523 \break
  \iCOLO
     University of Colorado,
     Boulder, Colorado 80309 \break
  \iCOL
     Columbia University,
     New York, New York 10027 \break
  \iFER
     INFN Sezione di Ferrara and Universit\`a di Ferrara,
     I-44100 Ferrara, Italy \break
  \iFRA
     INFN  Lab. Nazionali di Frascati,
     I-00044 Frascati, Italy \break
  \iILL
     University of Illinois,
     Urbana, Illinois 61801 \break
  \iLBL
     E.O. Lawrence Berkeley Laboratory, University of California,
     Berkeley, California 94720 \break
  \iMIT
     Massachusetts Institute of Technology,
     Cambridge, Massachusetts 02139 \break
  \iMASS
     University of Massachusetts,
     Amherst, Massachusetts 01003 \break
  \iMISS
     University of Mississippi,
     University, Mississippi  38677 \break
  \iMOSC
    Moscow State University,
    Institute of Nuclear Physics
    119899 Moscow, Russia    \break
  \iNAG
     Nagoya University,
     Chikusa-ku, Nagoya 464 Japan  \break
  \iOREG
     University of Oregon,
     Eugene, Oregon 97403 \break
  \iPAD
     INFN Sezione di Padova and Universit\`a di Padova,
     I-35100 Padova, Italy \break
  \iPERU
     INFN Sezione di Perugia and Universit\`a di Perugia,
     I-06100 Perugia, Italy \break
  \iPISA
     INFN Sezione di Pisa and Universit\`a di Pisa,
     I-56100 Pisa, Italy \break
  \iRUT
     Rutgers University,
     Piscataway, New Jersey 08855 \break
  \iRAL
     Rutherford Appleton Laboratory,
     Chilton, Didcot, Oxon OX11 0QX United Kingdom \break
  \iSOGANG
     Sogang University,
     Seoul, Korea \break
  \iSOONG
     Soongsil University,
     Seoul, Korea  156-743 \break
  \iSLAC
     Stanford Linear Accelerator Center, Stanford University,
     Stanford, California 94309 \break
  \iTENN
     University of Tennessee,
     Knoxville, Tennessee 37996 \break
  \iTOH
     Tohoku University,
     Sendai 980 Japan \break
  \iVAND
     Vanderbilt University,
     Nashville, Tennessee 37235 \break
  \iWASH
     University of Washington,
     Seattle, Washington 98195 \break
  \iWISC
     University of Wisconsin,
     Madison, Wisconsin 53706 \break
  \iYALE
     Yale University,
     New Haven, Connecticut 06511 \break
  \dead
     Deceased \break
  \andgen
     Also at the Universit\`a di Genova \break
  \andper
     Also at the Universit\`a di Perugia \break
\rm
%

\end{center}





\end{document}